\documentclass[12pt]{article}
\usepackage{amsmath,epsf,amssymb,latexsym,cite}
\usepackage[ps,dvips,matrix,arrow,frame,import,curve,color]{xy}

\newtheorem{theorem}{Theorem}

\newtheorem{lemma}{Lemma}

\newif\iffigs\figstrue

\DeclareFontFamily{U}{rsf}{}
\DeclareFontShape{U}{rsf}{m}{n}{
  <5> <6> rsfs5 <7> <8> <9> rsfs7 <10-> rsfs10}{}
\DeclareMathAlphabet\Scr{U}{rsf}{m}{n}

\def\pplogo{\vbox{\kern-\headheight\kern -29pt
\halign{##&##\hfil\cr&{%\sc
\ppnumber}\cr\rule{0pt}{2.5ex}&\ppdate\cr}
}}
\makeatletter
\def\ps@firstpage{\ps@empty \def\@oddhead{\hss\pplogo}%
  \let\@evenhead\@oddhead % in case an article starts on a left-hand page
}
\def\maketitle{\par
 \begingroup
 \def\thefootnote{\fnsymbol{footnote}}
 \def\@makefnmark{\hbox{$^{\@thefnmark}$\hss}}
 \if@twocolumn
 \twocolumn[\@maketitle]
 \else \newpage
 \global\@topnum\z@ \@maketitle \fi\thispagestyle{firstpage}\@thanks
 \endgroup
 \setcounter{footnote}{0}
 \let\maketitle\relax
 \let\@maketitle\relax
 \gdef\@thanks{}\gdef\@author{}\gdef\@title{}\let\thanks\relax}
\makeatother

\def\O{\Scr{O}}
\def\C{{\mathbb C}}
\def\P{{\mathbb P}}

\def\Z{{\mathbb Z}}

\def\Hom{\operatorname{Hom}}

\def\sHom{\operatorname{\Scr{H}\!\!\textit{om}}}
\def\sExt{\operatorname{\Scr{E}\!\textit{xt}\,}}
\def\Ext{\operatorname{Ext}}
\def\End{\operatorname{End}}

\def\Tr{\operatorname{Tr}}

\def\GU{\operatorname{U{}}}

\def\Ainf{{A_\infty}}
\def\id{{\mathbf{1}}}

\def\CY{Calabi--Yau}

\def\cA{{\Scr A}}
\def\cB{{\Scr B}}

\def\cI{{\Scr I}}

\def\cE{{\Scr E}}
\def\cF{{\Scr F}}

\def\DC{\mathbf{D}}

\def\ff#1#2{{\textstyle\frac{#1}{#2}}}

\def\mf#1{\mathfrak{#1}}

\def\ssd{{\mathbf{d}}}

\begin{document}
\setcounter{page}0
\def\ppnumber{\vbox{\baselineskip14pt
\hbox{SU-ITP-04/43}
\hbox{DUKE-CGTP-04-11}
\hbox{ILL-(TH)-04-7}
\hbox{hep-th/0412209}}}
\def\ppdate{December 2004} \date{}

\title{\LARGE Computation of Superpotentials for D-Branes\\[10mm]}
\author{
Paul S.~Aspinwall\\[2mm]
\normalsize Department of Physics and SLAC \\
\normalsize Stanford University \\
\normalsize Stanford, CA 94305/94309\\
\normalsize and\\
\normalsize Center for Geometry and Theoretical Physics \\
\normalsize Box 90318 \\
\normalsize Duke University \\
\normalsize Durham, NC 27708-0318\\[8mm]
Sheldon Katz\\[2mm]
\normalsize Departments of Mathematics and Physics\\
\normalsize University of Illinois at Urbana-Champaign\\
\normalsize Urbana, IL  61801 
}

{\hfuzz=10cm\maketitle}

\def\Large{\large}
\def\LARGE{\large\bf}

\vskip 1cm

\begin{abstract}
We present a general method for the computation of tree-level
superpotentials for the world-volume theory of B-type D-branes. This
includes quiver gauge theories in the case that the D-brane is
marginally stable. The technique involves analyzing the
$A_\infty$-structure inherent in the derived category of coherent
sheaves. This effectively gives a practical method of computing
correlation functions in holomorphic Chern--Simons theory. As an
example, we give a more rigorous proof of previous results concerning
3-branes on certain singularities including conifolds. We also provide
a new example.
\end{abstract}

\vfil\break

%%%%%%%%%%%%%%%%%%%%%%%%%%%%%%%%%%%%%%%%%%%%%%%%%%%%%%%%%%%%%%%%

\section{Introduction}    \label{s:intro}

Consider a type II superstring compactification on a \CY\ threefold
$X$. BPS D-branes that ``wrap cycles'' within $X$ and fill the noncompact
spacetime give rise to an effective $\mathcal{N}=1$, $d=4$ supersymmetric gauge
theory arising from the D-brane world-volume. As is well-known
\cite{W:DII}, if $N$ irreducible D-branes wrap the same cycle, one
obtains a model with $\GU(N)$ gauge symmetry.

One may obtain more general supersymmetric gauge theories in the form
of quiver gauge theories by considering marginally stable
D-branes. Suppose a given D-brane (which may consist of multiple
copies of some irreducible D-branes) is marginally stable with respect
to decay into $N_1$ copies of some (irreducible) D-brane plus $N_2$ copies of
another D-brane, etc., then one obtains a gauge theory with gauge
group $\GU(N_1)\times \GU(N_2)\times\ldots$. The fact that the given
D-brane is marginally stable means that there will be massless open
strings between the decay products. These strings give rise to
massless chiral supermultiplets in bifundamental $(\overline{\mathbf{N}}_1,
\mathbf{N_2})$ representations etc.

These $\mathcal{N}=1$, $d=4$ supersymmetric gauge theories will, in
general, have a nontrivial superpotential expressible as a function of
the chiral superfields. The purpose of this paper is to describe a
systematic and general method for the computation of this
superpotential at tree level directly from the algebraic geometry of
$X$.

There are two types of BPS D-branes on a \CY\ threefold --- the
so-called A-type and B-type. The A-type D-branes are described by
special Lagrangian cycles within $X$ \cite{BBS:5b} and are, in
principle, described completely by the language of the Fukaya category
\cite{Fuk:cat,Fuk:TQFT,FOOO:bar}. Having said that, the Fukaya
category is extremely difficult to deal with explicitly for any
example of a \CY\ threefold.

The other ``B-type'' D-branes are described by $\DC(X)$, the derived
category of coherent sheaves on $X$
\cite{Kon:mir,Doug:DC,AL:DC,me:TASI-D}. While, at first sight, the
derived category may appear to be mathematically formidable, it is
actually very useful for direct computations in any given example.

In this paper we therefore focus on the problem of computing the
superpotential for B-type D-branes. The easiest route for computing
superpotentials for A-branes in many situations is by reducing to the B-brane
case by using mirror symmetry.

Various methods leading to proposals for superpotentials in several
examples have already appeared in the literature
\cite{KW:coni,MP:AdS,CKV:quiv,CFIKV:,DGJT:D,FHH:toric,HLL:min}. Each
of these papers has used a somewhat indirect approach to analyzing the
superpotential. Here we will give a very direct method that can be
applied for any collection of B-type D-branes on any \CY\ manifold. It
uses similar ideas to those used in demonstrations of homological
mirror symmetry given in \cite{Polsh:Aie}.

In a recent paper \cite{ADED:obs} this problem was studied in the
Landau--Ginzburg phase of the B-model. It is believed that this should
give the same result as in the \CY\ phase, i.e., the case of interest
here. The only worry would be that, in current understanding, the
Landau--Ginzburg analysis appears to collapse the derived category
somewhat by identifying objects under a shift of 2. This might change
the results of computations of superpotentials in some cases.

It has long been known \cite{W:CS} that the information concerning the
superpotential is contained in a holomorphic Chern--Simons theory of
the \CY\ threefold in question. The propagator in this field theory
appears to require a complete knowledge of the metric of $X$ and so
cannot be computed in general. Here we recast the holomorphic
Chern--Simons theory in a form described purely by homological algebra
and algebraic geometry.

The logic of this argument is very similar to that used when arguing
that B-branes, which are originally described by Dolbeault cohomology,
are ultimately described by $\DC(X)$. Indeed, all we need do is to
supplement this argument by a product structure. The wedge product of
Dolbeault cohomology becomes a ``composition'' product in
$\DC(X)$. 

Having done this change of language to algebraic geometry one can then
use \v Cech cohomology and a knowledge of locally-free resolutions to
perform a computation of the superpotential. Our method applies, in
principle, to a computation involving any D-brane (i.e., any object in
the derived category) on any \CY\ threefold. The only obstacle in
general to the computation is the stamina required to compute \v Cech
cohomology when several patches are required, and dealing with
potentially long locally-free resolutions.

All the technical machinery of computing superpotentials is tied up in
the $\Ainf$-algebra language (see, for example,
\cite{GugSta:Ainf,PS:Ainf}). We therefore begin with a review of the
required facts in section \ref{s:Ainf}.

In section \ref{s:sup} we discuss general features of the
way that the superpotential is described by correlation functions in
the topological B-model. An interesting result is that, although one
can define a generalized superpotential on the ``thickened'' moduli
space of the topological field theory, it essentially contains no more
information than the physical superpotential. We also discuss the
uniqueness of the superpotential computed by the topological field theory.

In section \ref{s:CS} we show how holomorphic Chern--Simons theory can
be restated in terms more appropriate to algebraic geometry and in
section \ref{s:prac} we compute some examples. We are able to verify
some results concerning 3-branes on conifold singularities. We also
compute a new result based on a 5-brane wrapping a particular $\P^1$.

%%%%%%%%%%%%%%%%%%%%%%%%%%%%%%%%%%%%%%%%%%%%%%%%%%%%%%%%%%%%%%%%%%%

\section{$\Ainf$ Algebras and Categories}  \label{s:Ainf}

The superpotential in these $\mathcal{N}=1$ gauge theories is
intimately related to the structure of an $\Ainf$ algebra (in the case
of a single stable D-brane) or category (in the case of a
quiver). This has been discussed in
\cite{Laz:super,Toms:Ainf,DGJT:D,HLW:Ainf}. We begin, therefore, with a
review of $\Ainf$ algebras following \cite{GugSta:Ainf,PS:Ainf}.

Let $V$ be a vector space with a $\Z$-grading and let $T(V)$ be the
resulting graded tensor algebra
\begin{equation}
  T(V) = \bigoplus_{n=1}^\infty V^{\otimes n}.
\end{equation}
If $a\in V$, we will denote the grade of $a$ by $|a|$. By the usual
abuse of notation we will often write $(-1)^a$ rather than $(-1)^{|a|}$. If
$f$ and $g$ are operators of given degrees, we use the rule
\begin{equation}
  (f\otimes g)(a\otimes b) = (-1)^{|g|.|a|}f(a)\otimes g(b).  \label{eq:sign}
\end{equation}

Now let $\ssd$ be a derivative with degree 1, with respect to the
grading, acting on $T(V)$ obeying
the graded Leibniz rule
\begin{equation}
  \ssd(a\otimes b) = \ssd(a)\otimes b + (-1)^a a\otimes\ssd(b).
     \label{eq:Leib}
\end{equation}
We also demand
\begin{equation}
  \ssd^2=0.  \label{eq:d2=0}
\end{equation}
The Leibniz rule (\ref{eq:Leib}) means that $\ssd$ is entirely determined
by its restriction to $V$. Let us denote this restriction as
$(\ssd)_V$. One can then decompose
\begin{equation}
  (\ssd)_V = d_1+d_2+\ldots,
\end{equation}
where
\begin{equation}
  d_k: V \to V^{\otimes k}.
\end{equation}

Let $V[1]$ denote the vector space $V$ with all the grades decreased
by one and let $s:V\to V[1]$ be the obvious map of degree $-1$. We can
now define our $\Ainf$ algebra $A$:
\begin{equation}
  A = (V[1])^*.
\end{equation}
together with its higher products
\begin{equation}
  m_k: A^{\otimes k}\to A,
\end{equation}
given by the dual of $s^{\otimes k}\cdot d_k\cdot s^{-1}$. The map $m_k$
thus has degree $2-k$.\footnote{The grades of vector spaces are
negated upon dualizing. Thus, when a map between vector spaces is
dualized, its direction is reversed but its degree remains the same.}

The condition (\ref{eq:d2=0}) then becomes equivalent to \cite{Kel:Ainf}
\begin{equation}
  \sum_{r+s+t=n} (-1)^{r+st}m_u(\id^{\otimes r}\otimes m_s\otimes \id^{\otimes
  t})=0,   \label{eq:Ainf}
\end{equation}
for any $n>0$, where $u=n+1-s$. One may view (\ref{eq:Ainf}) as the defining
relations for an $\Ainf$ algebra.

It is easy to extend the idea of an $\Ainf$-algebra to an
$\Ainf$-category \cite{KS:mir}. Such a category consists of objects
and morphisms in the usual way except that morphisms, under $k$-fold
compositions, satisfy the relations (\ref{eq:Ainf}). In particular, an
$\Ainf$-category need not be a category in the usual sense since
composition of morphisms need not be associative.

Now suppose we have another graded vector space $U$ with its own
differential $\ssd$ acting on $T(U)$. It is natural to consider maps
$g:T(U)\to T(V)$ which commute with $\ssd$. We impose the condition
\begin{equation}
  g(a\otimes b) = (-1)^{|g||a|}g(a)\otimes g(b),
\end{equation}
so that such maps are defined completely by their restriction to $U$.

If $B$ is the $\Ainf$-algebra constructed from $U$, such a $g$ gives
rise to an ``$\Ainf$-morphism'' given by maps
\begin{equation}
  f_k:A^{\otimes k} \to B,
\end{equation}
constructed in the obvious way from $g$ above. The condition that $g$
commutes with $\ssd$ then becomes
\begin{equation}
  \sum_{r+s+t=n}(-1)^{r+st} f_u(\id^{\otimes r}\otimes m_s\otimes \id^{\otimes
  t}) =
  \sum_{\genfrac{}{}{0pt}{}{1\leq r\leq n}{i_1+\ldots+i_r=n}}
     \!\!(-1)^q m_r(f_{i_1}\otimes
  f_{i_2}\otimes\cdots\otimes f_{i_r}), \label{eq:Amorph}
\end{equation}
for any $n>0$ and $u=n+1-s$ again. The sign on the right is given by
\begin{equation}
  q = (r-1)(i_1-1) + (r-2)(i_2-1) + \ldots + (i_{r-1}-1).
\end{equation}

Note that $m_1:A\to A$ is a degree one map satisfying $m_1\cdot
m_1=0$. It thus gives $A$ the structure of a graded differential
complex, and we may take cohomology to yield $H^*(A)$. By choosing
representatives of each cohomology class we may define an embedding
\begin{equation}
  i : H^*(A) \hookrightarrow A.
\end{equation}
Thanks to a theorem by Kadeishvili \cite{Kad:Ainf}, we may define an
$\Ainf$ structure on $H^*(A)$ such that
\begin{enumerate}
\item There is an $\Ainf$ morphism $f$ from $H^*(A)$ to $A$ with
$f_1$ equal to the embedding $i$.
\item $m_1=0$.
\end{enumerate}
Here, $m_1$ refers to the $\Ainf$ structure on $H^*(A)$.
This $\Ainf$ structure is not unique, but it is unique up
to $\Ainf$-isomorphisms, as will be discussed in a more general situation
in Lemma~\ref{lem:ainfiso} below.  An $\Ainf$-algebra with $m_1=0$ is called
a {\em minimal\/} $\Ainf$-algebra.

It is quite easy to construct Kadeishvili's $\Ainf$-structure in
practice.
A rather simple example of an $\Ainf$-algebra is given by $m_k=0$ for
$k\geq3$. Such an algebra is called a {\em differential graded
algebra}, or dga. In this paper, we will need to put an $\Ainf$ structure on
the cohomology of a dga, which may be done explicitly as follows. Let
$m_1$ on $A$ be denoted $d$, and let $m_2(a\otimes b)$ be denoted $a\cdot
b$.

Putting $n=2$ in (\ref{eq:Amorph}), and using the fact that $m_1=0$ in
$H^*(A)$, yields
\begin{equation}
\label{eq:m2}
  im_2 = (i\cdot i) + df_2. 
\end{equation}
Since $d(i\cdot i)=0$, we must
define $m_2$ on $H^*(A)$ as the cohomology class of $i\cdot i$.  We may also 
use this to define a choice of
$f_2:H^*(A)^{\otimes2}\to A$ (up to an element in the kernel of
$d$). Next, putting $n=3$ in (\ref{eq:Amorph}) yields\footnote{There
appears to be a typo in \cite{Kel:Ainf} discarding too many terms.}
\begin{equation}
  im_3 = f_2(\id\otimes m_2)-f_2(m_2\otimes\id) + (i\cdot f_2)
  -(f_2\cdot i) + d f_3. \label{eq:m3}
\end{equation}
Direct computation using (\ref{eq:m2})
shows that $d(f_2(\id\otimes m_2)-f_2(m_2\otimes\id) + 
(i\cdot f_2)  -(f_2\cdot i))=0$, so as before 
this defines $m_3$ and allows us to choose a definition for
$f_3$. Clearly this process continues and defines all the products for
the $\Ainf$ algebra on $H^*(A)$.

This construction of the $\Ainf$ algebra on $H^*(A)$ may be rephrased
following \cite{Merk:Dol,KS:mir} in a language which will also be
useful to us. Suppose we define a projection $p:A\to H^*(A)$ such that
$p\circ i=1$ and furthermore assume that we have a map $H:A\to A$ of
degree $-1$ such that $1-i\circ p = dH + Hd$.

Clearly, $m_2$ is defined as $p\circ(i\cdot i)$ as before. For
$k>2$, we then define
\begin{equation}
  m_k = \sum_T \pm m_{k,T},
\end{equation}
where the sum is over all trees $T$ with $k$ branch tips at the top
and one root. These trees look like Feynman diagrams of a $\phi^3$
field theory and are computed accordingly, with $m_2$ of $A$ acting as the
cubic coupling and $H$ acting as the propagator. We refer to
\cite{KS:mir} for more details.

\medskip
We say that two dga's $A$ and $B$ are {\em quasi-isomorphic\/} if there
is a homomorphism of dga's $g:A\to B$ (i.e.\ preserving the respective
products and commuting with the differentials) inducing an isomorphism
on the respective cohomologies, which can then be identified.
A simple extension of the ``uniqueness up to $\Ainf$-isomorphism'' part of
the above construction proves the following.

\begin{lemma}
\label{lem:ainfiso}
  Suppose that $A$ and $B$ are quasi-isomorphic dga's, determining
$\Ainf$ structures on $H^*(A)\simeq H^*(B)$ as above.  Then these two
$\Ainf$ algebras are $\Ainf$-isomorphic.
\end{lemma}

\smallskip
Before describing the simple proof, we recall that
$\Ainf$-quasi-isomorphisms are simply $\Ainf$-morphisms which are
quasi-isomorphisms, i.e.\ which induce isomorphisms between the
respective $m_1$-cohomologies.  We will need the result that
$\Ainf$-quasi-isomorphisms have homotopy inverses, see \cite{Kel:Ainf}
and the references therein.  We also need the result that an $\Ainf$-morphism
$f$ between minimal $\Ainf$-algebras is an isomorphism if and only if
$f_1$ is an isomorphism.
 
\smallskip
Let $f$ be an $\Ainf$-quasi-isomorphism from $H^*(A)$ to $A$ and
$g$ be an $\Ainf$-quasi-isomorphism from $H^*(B)$ to $B$ as in
Kadeishvili's theorem.  Let $\phi:A\to B$ be the given
quasi-isomorphism of dga's, which, viewing $A$ and $B$ as
$\Ainf$-algebras can be viewed as describing an
$\Ainf$-quasi-isomorphism.  Let $h$ be a homotopy inverse of $g$.
Then $r=h\circ\phi\circ f$ is an $\Ainf$-morphism from $H^*(A)$ to
$H^*(B)$.  Here $\circ$ denotes the composition of $\Ainf$-morphisms,
see e.g.\ \cite{Kel:Ainf}.  But $r_1:H^*(A)\to H^*(B)$ is an
isomorphism by definition since $\phi$ is a quasi-isomorphism, while
$H^*(A)$ and $H^*(B)$ are minimal $\Ainf$-algebras.  Hence $r$ is an
isomorphism by the discussion above.

%%%%%%%%%%%%%%%%%%%%%%%%%%%%%%%%%%%%%%%%%%%%%%%%%%%%%%%%%%%%%%%%%%%

\section{D-Branes and Superpotentials} \label{s:sup}

To begin with, assume we have a D-brane that consists of a vector
bundle $E\to X$. This was the case studied by Witten in
\cite{W:CS}. The open strings in the B-model correspond to elements of 
Dolbeault cohomology $H^{0,q}_{\bar\partial}(X,\End(E))$. 

The vertex operators corresponding to $q=0$ yield massless vector
bosons in the uncompactified 4 dimensions which give rise to a gauge
theory. The vector bundle $E$ is said to be {\em simple\/} if $\Hom(E,E)
=\End(E)=\C$. In this case we have one vector boson and the gauge
group is $\GU(1)$. Similarly if $E$ is $(E_0)^{\oplus N}$, where $E_0$
is simple, then the gauge group is $\GU(N)$.

The vertex operators corresponding to $q=1$ yield massless scalars and
fermions in four dimensions coming from chiral supermultiplets. These
transform in the adjoint representation of $\GU(N)$.  Let $A$ denote
the Hilbert space of open string states.  The effective D-brane
world-volume theory contains a superpotential which is a holomorphic
function of these chiral superfields.

In \cite{BDLR:Dq} it was shown that this superpotential could be
computed in terms of correlation functions in the associated {\em
topological\/} quantum field theory. The result is as follows.

The open strings are associated to local vertex operators $\psi_i$ in
the topological field theory. These $\psi_i$'s may be viewed as a
basis for $A$. To each such vertex operator, one may construct a
1-form operator 
\begin{equation}
  \psi_i^{(1)} = \ff1{\sqrt2}\left\{
    G^-_{-\frac12} + \overline{G}^-_{-\frac12},\psi_i\right\},  \label{eq:desc}
\end{equation}
These 1-form operators may be used to deform the topological field
theory (at least to first order):
\begin{equation}
  S \to S + \sum_i Z_i\psi_i^{(1)},  \label{eq:Sdef}
\end{equation}
where the $Z_i$ are complex numbers as far as the topological field
theory is concerned. The $Z_i$ are (the scalar components of) chiral
superfields in the effective world-volume theory. The deformations
(\ref{eq:Sdef}) correspond to giving vacuum expectations values to
these fields. Thus, {\em the chiral superfields are naturally dual to
  the vertex operators of the topological quantum field theory}.

Let $q_i\in\Z$ denote the ghost number of $\psi_i$. Then
$\psi_i^{(1)}$ has ghost number $q_i-1$. The only operators that can
be used to deform the untwisted conformal field theory associated to
the D-brane must have $q_i=1$. We would like to extend our discussion
to the ``thickened'' moduli space of \cite{W:AB}. In this picture, all
ghost numbers are allowed. This gives rise to a generalized space of
chiral superfields where $Z_i$ has a grade $1-q_i$.  We also have a
generalized superpotential $\mathbf{W}$ which is a function of all the
$Z_i$'s. Always remember, though, that only the fields of grade zero
are true chiral superfields.

Following the conventions of \cite{HLW:Ainf}, we define correlation
functions for $k+1$ open string vertex operators:
\begin{equation}
  B_{i_0,i_1,\ldots,i_{k}} = (-1)^{\zeta_1+\zeta_2+\ldots+\zeta_{k-1}}
    \langle \psi_{i_0}\,\psi_{i_1}\,P\int\psi_{i_2}^{(1)}\,
    \int\psi_{i_3}^{(1)}\ldots\int\psi_{i_{k-1}}^{(1)}\,
    \psi_{i_k}\rangle,  \label{eq:Bdef}
\end{equation}
where we introduce the notation $\zeta_j=1-q_{i_j}$. The integrals in
this correlation function are over segments of the boundary so as to
preserve the path ordering. A choice of regulator needs to be made in
order to fully define these correlation functions as was done in
\cite{HLW:Ainf}. We will avoid making such a choice, giving rise to
ambiguities which we discuss at the end of this section.

It was shown in \cite{HLW:Ainf} that these correlators satisfy the
following cyclicity property
\begin{equation}
   B_{i_0,i_1,\ldots,i_{k}} =
    (-1)^{\zeta_k(\zeta_0+\zeta_1+\ldots+\zeta_{k-1})}
    B_{i_k,i_0,i_1,\ldots,i_{k-1}}. \label{eq:cyc}
\end{equation}
In the case of $N$ copies of a simple D-brane, the fields $Z_i$
naturally form $N\times N$ matrices. We may now write the superpotential
\begin{equation}
  \mathbf{W} = \Tr\left(\sum_{k=2}^\infty\,\sum_{i_0,i_1,\ldots,i_{k}}\!
    \frac{B_{i_0,i_1,\ldots,i_{k}}}{k+1}Z_{i_0}Z_{i_1}\ldots Z_{i_{k}}\right).
    \label{eq:Wdef}
\end{equation}
Note that this trace has a graded cyclicity property consistent with
(\ref{eq:cyc}).

The correlation functions of a topological quantum field theory are
subject to various constraints due to sewing conditions as discussed
in \cite{Dub:2d,Segal:func} and, in particular, \cite{Laz:oc}. The
open string ``pair of pants'' diagram associates a bilinear product of
degree 0 to $A$. Anticipating the connection with $\Ainf$ algebras, we
denote this
\begin{equation}
  m_2:A\otimes A \to A.
\end{equation}
If $X$ is a \CY\ threefold, there is also a ``trace map'' of degree $-3$
\begin{equation}
  \gamma:A\to\C.  \label{eq:trmap}
\end{equation}
It follows that our desired correlation function may be written in the
form
\begin{equation}
  B_{i_0,i_1,\ldots,i_{k}} =
    \gamma\left(m_2\left(m_k(\psi_{i_0},\psi_{i_1},\ldots,
     \psi_{i_{k-1}}),\psi_{i_k}\right)\right),
\end{equation}
for maps of degree $2-k$
\begin{equation}
  m_k:A^{\otimes k} \to A.
\end{equation}
It was shown in \cite{HLW:Ainf} that these products do indeed obey the
conditions (\ref{eq:Ainf}) and thus give $A$ the structure of an
$\Ainf$ algebra.

Comparing this structure with the description of $\Ainf$ algebras in
section \ref{s:Ainf}, it should be clear that the chiral superfields
$Z_i$ play the role of generators of the space $V$. The shift by one
comes from (\ref{eq:desc}) and the dualizing comes from (\ref{eq:Sdef}).
Since the structure of the $\Ainf$ algebra is simpler to describe in
terms of $T(V)$, it should be enlightening to rephrase the above in
this language.

The degree $-3$ pairing $\gamma(m_2(-,-))$ is non-degenerate on $A$
and simply corresponds to Serre duality. It naturally dualizes to
produce a map
\begin{equation}
  \eta:\C \to V\otimes V,
\end{equation}
of degree $-1$.

If $Z_i$ is a homogeneous basis for $V$,
\begin{equation}
  \eta(1) = \sum_i Z_i\otimes\hat Z_i,
\end{equation}
where $\hat Z_i$ are viewed as the ``Serre dual'' of $Z_i$.

We write a basis of $V$ as follows. Let $X_1\ldots X_n$ be a basis of
the degree 0 part of $V$. These are therefore the true chiral
superfields in the four-dimensional theory.  Assume, for now, that $E$ is
simple, i.e., $\Hom(E,E)=\C$.  In other words, there is a unique
``identity'' vertex operator for open strings beginning and ending on
$E$. This is dual to an element denoted $e\in V$ of degree 1. Serre
duality can now be used to give a basis $\hat X_i$ of the degree $-1$
part of $V$ and a generator $\hat e$ of the degree $-2$ part of $V$,
where
\begin{equation}
  \eta(1) = e\otimes\hat e + \hat e \otimes e +
    \sum_\alpha X_\alpha\otimes\hat X_\alpha + \sum_\alpha X_\alpha
    \otimes\hat X_\alpha. \label{eq:eta1} 
\end{equation}

Viewing the superpotential $\mathbf{W}$ as an element of $T(V)$ and using
(\ref{eq:Wdef}), the higher products of the $\Ainf$ algebra can be
rephrased in the language of section \ref{s:Ainf} as the beautifully
simple statement 
\begin{equation}
  \ssd \hat Z_i = \frac{\partial\mathbf{W}}{\partial Z_i}. \label{eq:dZ}
\end{equation}
Since $T(V)$ is the {\em non-commutative\/} algebra generated by $Z_i$,
some care is needed in defining the partial derivative in (\ref{eq:dZ}).
The recipe is as follows. The cyclic trace property (\ref{eq:cyc})
allows $\mathbf{W}$ to be written with any of the generators at the
front. $\partial\mathbf{W}/\partial Z_i$ is then defined as the sum of all the
possible forms of $W$ under the trace property with $Z_i$ at the
front, with said $Z_i$ removed. Clearly this coincides with the
usual definition of derivative in commutative algebra.

The identity vertex operator has special properties under the higher
products as shown in \cite{HLW:Ainf}. Let $\psi_0$ be the identity
operator. Then
\begin{equation}
\begin{split}
  m_2(\psi_0,\psi_i) = m_2(\psi_i,\psi_0) &= \psi_i\\
  m_k(\psi_{i_1},\psi_{i_2},\ldots,\psi_0,\ldots) &= 0, 
     \qquad\hbox{for $k>2$.}
\end{split}
\end{equation}
Carefully computing signs, it follows from (\ref{eq:dZ}) that
\begin{equation}
  \mathbf{W} =\frac12\sum_i\left((-1)^{Z_i}\hat Z_i\otimes e\otimes Z_i - 
      \hat Z_i\otimes Z_i\otimes e\right) + 
    \hbox{terms not containing $e$}.  \label{eq:W1}
\end{equation}
In (\ref{eq:W1}) we have also dropped terms which can be deduced from the
cyclicity property (\ref{eq:cyc}).

So far we have discussed one simple D-brane. It is very easy to
generalize to the case of a collection of D-branes
\begin{equation}
  E_1^{\oplus N_1} \oplus E_2^{\oplus N_2}\oplus\ldots,
\end{equation}
forming a $\GU(N_1)\times\GU(N_2)\times\ldots$ quiver gauge theory,
where each $E_j$ is simple. In order to form a quiver gauge theory
free from tachyons or peculiar vector bosons we are required to impose
\cite{AM:delP,Herz:ouch}
\begin{equation}
  \Hom(E_j,E_k) = 0, \qquad\hbox{for $j\neq k$}.  \label{eq:noHom}
\end{equation}
This means that the only degree zero vertex operators in $A$ remain
multiples of identity maps $E_j\to E_j$.

The effect of passing to a quiver gauge theory is that we must now
think in terms of $\Ainf$ categories rather than algebras.  This
amounts to little more than bookkeeping as follows.  The elements of
$A$ should be viewed as morphisms between D-branes and, as such, as
elements of $H^{0,*}_{\bar\partial}(X,\Hom(E_i,E_j))$. All we need do is
to rewrite (\ref{eq:W1}) as
\begin{equation}
  \mathbf{W} = \frac12\Tr\left(\sum_i (-1)^{Z_i}\hat Z_i\otimes e\otimes Z_i - 
      \sum_i \hat Z_i\otimes Z_i\otimes e + 
    \hbox{terms not containing $e$}\right),  \label{eq:W2}
\end{equation}
where now $Z_i$ are matrices. The ``$\otimes$'' in (\ref{eq:W2}) now
implicitly includes matrix multiplication and the concept of
composition of morphisms between different objects. The symbol $e$ now
refers to a square $N_j\times N_j$ matrix with entries dual to the
identity operator of a given simple D-brane $E_j$. The composition of
morphisms implied by the superpotential must begin and end on the same
D-brane so that a trace may then be taken. This is equivalent to the
statement that the superpotential is gauge invariant.

The equation (\ref{eq:dZ}) remains valid for the quiver theory. One
implicitly removes the trace and then na\"\i vely applies the rule for
differentiation we described above.

The general form of the superpotential can be further constrained. The
only vertex operators appearing have grade 0, 1, 2 or 3. Thus, the
$Z_i$ have grade 1, 0, $-1$ or $-2$, with the $e$ the only generator
with grade 1. Now, $\ssd e$ is of degree 2 and so must be a sum of
terms in $T(V)$ each with at least one $e$. The property of the
identity element of an $\Ainf$ algebra thus implies
\begin{equation}
  \ssd e = -e\otimes e.
\end{equation}
Similarly, since $X_\alpha$ has degree 0, we must have
\begin{equation}
  \ssd X_\alpha = X_\alpha\otimes e - e \otimes X_\alpha.
\end{equation}
Since $\hat X_\alpha$ is of degree $-1$,
\begin{equation}
  \ssd \hat X_\alpha = F(X_\beta) - \hat X_\alpha\otimes e -
      e\otimes\hat X_\alpha,
\end{equation}
where $F(X_\beta)$ is an arbitrary function of the
$X_\beta$'s. Finally $d\hat e =\partial\mathbf{W}/\partial e$ and is
completely determined by (\ref{eq:W2}). The result is that 
\begin{equation}
  \mathbf{W} = \Tr\left(W(X_\alpha) - \hat e \otimes e \otimes e
   +\sum_\alpha\left(\hat X_\alpha\otimes X_\alpha\otimes e -
   \hat X_\alpha\otimes e\otimes X_\alpha\right)\right),  \label{eq:Wgen}
\end{equation}
where $W(X_\alpha)$ is a completely arbitrary function of all the
chiral superfields $X_\alpha$. This function
is, of course, the physical superpotential.

For any $Z_i\in V$, using (\ref{eq:dZ}) and (\ref{eq:Wgen}) it is a
simple matter to show
\begin{equation}
  \ssd^2 Z_i=0.
\end{equation}
There are therefore three remarkable properties of (\ref{eq:Wgen}):
\begin{enumerate}
\item The $\Ainf$ relations are trivially satisfied. There is no need
  to go through the computation of \cite{HLW:Ainf}.
\item The generalized superpotential associated to the thickened
  moduli space is determined completely by $W(X_\alpha)$ --- the
  physical superpotential on the physical moduli space.
\item The $\Ainf$ relations are satisfied for completely arbitrary
  $W(X_\alpha)$. 
\end{enumerate}

We should perhaps point out that much of the simplification we have
found here is due to the constraint (\ref{eq:noHom}) for a physical
quiver. Had we not imposed this, we could not have used the special
properties of the identity operator.

The $\Ainf$-morphisms are also simplified when passing to the dual
language of the chiral superfields. An $\Ainf$-morphism from a theory
with superfields $Z_i$ to a theory with superfields $Y_\alpha$ is
simply an analytic map
\begin{equation}
  Y_\alpha = g_\alpha(Z_1,Z_2,\ldots).
\end{equation}
The complicated expression (\ref{eq:Amorph}) is restated as $g$
commuting with $\mathbf{d}$. If any $f_k$ is nonzero for $k\geq2$,
this map of superfields is nonlinear.

We will be using Kadeishvili's theorem of section \ref{s:Ainf} to
compute the desired $\Ainf$-structure yielding the superpotential, It
is important to note that this theory only gives this structure {\em
up to an $\Ainf$-isomorphism.} It follows that we will only be able to
determine the superpotential {\em up to a nonlinear change in
superfields\/} where this nonlinear map is invertible and commutes
with $\mathbf{d}$.

It is not surprising that there is an ambiguity in the
superpotential. From the four-dimensional field theory point of view,
the topological B-model knows nothing about the kinetic term and so
one is free to apply nonlinear redefinitions to the chiral
superfields. From the point of the view of the string worldsheet,
contact terms arise from the vertex insertion point coalescing at the
ends of the integration regions. Such contact terms are known to
introduce ambiguities as in \cite{Kut:cnt}.

That these ambiguities exist is therefore not a surprise, but we have a
very precise form of the ambiguity --- the nonlinear redefinition of 
the superfields must commute with $\mathbf{d}$. It would be
interesting to find the physics behind this statement but we will not
attempt to pursue this question here.

%%%%%%%%%%%%%%%%%%%%%%%%%%%%%%%%%%%%%%%%%%%%%%%%%%%%%%%%%%%%%%%%%%%

\section{Holomorphic Chern--Simons Theory}  \label{s:CS}

\def\fa{\mathsf{A}}
In \cite{W:CS}, it was shown how to  exactly compute the correlation
functions (\ref{eq:Bdef}), at least for one D-brane $E$ and for vertex
operators in $H^{0,1}_{\bar\partial}(X,\End(E))$. One defines a {\em
holomorphic Chern--Simons theory\/} with action
\begin{equation}
S = \int_X \Tr\left(\fa\wedge \bar\partial\fa + \ff23\fa\wedge\fa\wedge
\fa\right)\wedge\Omega,  \label{eq:hCS}
\end{equation}
where the field $\fa$ is a $(0,1)$-form on $X$ taking values in
$\End(E)$, and $\Omega$ is a holomorphic $(3,0)$-form on $X$.

From this, the correlation functions are then computed as follows in
the language of section \ref{s:sup}. Let $A$ be the Hilbert space
$H^{0,*}(X,\End(E))$. The trace map (\ref{eq:trmap}) is given by
\begin{equation}
  \gamma(a) = \int_X \Tr(a)\wedge\Omega,
\end{equation}
while
\begin{equation}
  m_2(a,b) = a\wedge b,
\end{equation}
where composition in $\End(E)$ is implicit.
The computation of $m_k$ is then exactly as described by the tree
construction at the end of section \ref{s:Ainf}. The propagator is, of
course, the propagator of (\ref{eq:hCS}) which is given by
$H=G\bar\partial^\dagger$, where $G$ is the Green's operator inverting
the Laplacian. Since, for any differential form $\alpha$, \cite{Merk:Dol}
\begin{equation}
  \alpha = [\alpha]_{\textrm{Harm}} + \bar\partial G
  \bar\partial^\dagger + G\bar\partial^\dagger\bar\partial,
\end{equation}
we have the following (which was also effectively noted in \cite{DGJT:D}):
\begin{theorem}
The correlation functions in the holomorphic Chern--Simons theory are
associated with the $\Ainf$ algebra as computed in section \ref{s:Ainf},
where the dga is given by the Dolbeault complex of $\End(E)$-valued
$(0,q)$-forms together with the wedge product. The embedding, $i$, of
$H^{0,*}(X,\End(E))$ into this complex is given by Harmonic forms.
\end{theorem}

This formulation of holomorphic Chern--Simons theory is all very well
but it is not very practical. Computing the propagator
$G\bar\partial^\dagger$ would appear to require a knowledge of the
metric on $X$. Naturally this is not in the spirit of the topological
field theory. One general expects all computations in the topological
B-model to be cast in the language of algebraic geometry and thus not
require detailed knowledge of $X$, such as its metric.

The derived category program for B-branes \cite{Doug:DC} precisely
does this translation to algebraic geometry as reviewed in
\cite{me:TASI-D}. We need to extend this argument to include product
structures. What we will arrive at is an $\Ainf$-structure implicit in
the derived category that has been discussed in
\cite{Polsh:high,Polsh:Aie,Pol:AinfP}. Indeed, the equivalence we
derive in this section was also described in these references.

The key idea is that we have three natural dga's associated to three
different cohomologies, all of which may equally be used to analyze
the problem at hand. Suppose we have a holomorphic vector bundle $B$
with a product $\mu:B\otimes B\to B$. Let $\cB$ be the locally-free
sheaf of sections of $B$. In the case of interest, we will want
$B=\End(E)$, and $\cB=\sHom(\cE,\cE)$ with the product $\mu$ being given by
composition.  The useful dga's are then:
\begin{enumerate}
\item The Dolbeault complex of $(0,q)$-forms valued in $B$:
\begin{equation}
\xymatrix@1{
\ldots\ar[r]^-{\bar\partial}&\Gamma(\cA^{0,q-1}\otimes \cB)\ar[r]^-{\bar\partial}&
\Gamma(\cA^{0,q-1}\otimes \cB)\ar[r]^-{\bar\partial}&
\Gamma(\cA^{0,q-1}\otimes \cB)\ar[r]^-{\bar\partial}&\ldots,
}
\end{equation}
where $\Gamma$ denotes global section and $\cA^{0,q}$ is the sheaf of
$C^\infty$ $(0,q)$-forms on $X$. This yields Dolbeault cohomology
groups $H^*_{\bar\partial}(X,B)$. The product is given by the wedge
product combined with $\mu$. Putting $B=\End(E)$, this is the
description Witten originally used to formulate the B-model
\cite{W:CS}.
\item The \v Cech complex of \v Cech cochains associated to an open
  cover $\mf{U}$ for the locally-free sheaf $\cB$ of sections of $B$:
\begin{equation}
\label{eq:cechb}
\xymatrix@1{
\ldots\ar[r]^-{\delta}&\check{C}^{n-1}(\mf{U},\cB)
\ar[r]^-{\delta}&\check{C}^{n}(\mf{U},\cB)\ar[r]^-{\delta}&
   \check{C}^{n+1}(\mf{U},\cB)\ar[r]^-{\delta}&\ldots
}
\end{equation}
For sufficiently fine $\mf{U}$, the cohomology of this complex yields
the \v Cech cohomology groups $\check{H}^*(X,\cB)$. The product given
by the cup product combined with $\mu$ yields the dga.
\item Given an injective resolution of $\cB$:
\begin{equation}
\xymatrix@1{
0\ar[r]&\cB\ar[r]&\cI^0\ar[r]^-{i_0}&\cI^1\ar[r]^-{i_1}&\cI^2\ar[r]^-{i_2}
&\ldots,
} \label{eq:Ires}
\end{equation}
we may apply the global section functor, $\Gamma$, to yield a complex
\begin{equation}
\xymatrix@1@C=15mm{
\ldots\ar[r]^-{\Gamma(i_{n-2})}&\Gamma(\cI^{n-1})
\ar[r]^-{\Gamma(i_{n-1})}&\Gamma(\cI^{n})\ar[r]^-{\Gamma(i_{n})}&
      \Gamma(\cI^{n+1})\ar[r]^-{\Gamma(i_{n+1})}&\ldots, 
} \label{eq:Rg}
\end{equation}
whose cohomology yields the sheaf cohomology groups $H^*(X,\cB)$. The
resolution (\ref{eq:Ires}) extends $\mu$ naturally to a product:
\begin{equation}
  \mu:\cI^p\otimes\cI^q \to \cI^{p+q},
\end{equation}
which gives a dga structure to (\ref{eq:Rg}).
\end{enumerate}

There is a standard spectral sequence argument, as reviewed in
\cite{me:TASI-D} which shows that these three theories of cohomology
are equivalent. For example, one may define the double complex
\begin{equation}
\label{eq:epq}
  E_0^{p,q} = \check{C}^p(\mf{U},\cB\otimes\cA^{0,q}).
\end{equation}
To this we associate a single complex
\begin{equation}
  E^n = \bigoplus_{p+q=n} E_0^{p,q},
\end{equation}
with differential ${d}=\delta + (-1)^p\bar\partial$.  The $d$-cohomology
of $E^\bullet$ can be realized as the abutment of either of two
spectral sequences.  The first spectral sequence has $E_1$-term obtained 
from the $p$-cohomology of (\ref{eq:epq}):
\begin{equation}
\label{eq:e1term}
E_1^{p,q}=\check{H}^p(\cB\otimes\cA^{0,q}),
\end{equation}
and the second spectral sequence has $E_1$-term obtained 
from the $q$-cohomology of (\ref{eq:epq}).

Since the $\cA^{0,q}$ are fine sheaves (i.e.\ admit partitions of
unity), so are the $\cB\otimes\cA^{0,q}$. It follows that their higher
cohomologies vanish and (\ref{eq:e1term}) reduces to
\begin{equation}
E_1^{0,q}=\Gamma(\cB\otimes\cA^{0,q})
\end{equation}
with $E_1^{p,q}=0$ for $p>0$.  Here $\Lambda^{0,q}$ is the ring of
global $(0,q)$-forms on $X$.  This spectral sequence therefore
degenerates at $E_2$ and the $d$-cohomology of $E^\bullet$ is
isomorphic to the cohomology of $E_1^{0,\bullet}$.  In other words,
the chain map

\begin{equation}
\xymatrix@C=8mm{
\ldots\ar[r]^-{\bar\partial}&\Gamma(\cB\otimes\cA^{0,n-1})
\ar[r]^-{\bar\partial}\ar[d]^-{\xi}
&\Gamma(\cB\otimes\cA^{0,n})\ar[r]^-{\bar\partial}\ar[d]^-{\xi}&
   \Gamma(\cB\otimes\cA^{0,n+1})\ar[r]^-
{\bar\partial}\ar[d]^-{\xi}&\ldots\\
\ldots\ar[r]^-d&E^{n-1}\ar[r]^-d&E^{n}\ar[r]^-d&E^{n+1}\ar[r]^-d&\ldots
}
\end{equation}
is a {\em quasi-isomorphism\/}.  That is, $\xi$
induces an isomorphism between the cohomology of the two complexes.
Here, the $\xi$ are the natural maps $\Gamma(\cB\otimes\cA^{0,k})\to
\check{C}^0(\mf{U},\cB\otimes\cA^{0,k})\subset E^k$ expressing a global
section in terms of the given open cover.

It is easy to see that $\xi$ preserves the product structure between
the two complexes too. Thus, $\xi$ is a quasi-isomorphism of
dga's.

Turning to the other spectral sequence, the fact that 
\begin{equation}
\xymatrix@1{
  0\ar[r]&\O\ar[r]^-{\varepsilon}&\cA^{0,0}\ar[r]^-{\bar\partial}&
  \cA^{0,1}\ar[r]^-{\bar\partial}&\cA^{0,2}\ar[r]^-{\bar\partial}&\ldots,
}
\end{equation}
is exact (and remains exact upon tensoring with $\cB$) for
intersections in a suitably-chosen $\mf{U}$ means that
the $q$-cohomology of (\ref{eq:epq}) is 0 unless $q=0$, in which case
we simply get $\check{C}^p(\mf{U},\cB)$.  Thus this spectral sequence
degenerates as well, and the cohomology of $\check{C}^\bullet(\mf{U},\cB)$
coincides with the $d$-cohomology of $E^\bullet$.  More precisely, 
the chain map
\begin{equation}
\xymatrix@C=10mm{
\ldots\ar[r]^-{\delta}&\check{C}^{n-1}(\mf{U},\cB)
\ar[r]^-{\delta}\ar[d]^-{\varepsilon}
&\check{C}^{n}(\mf{U},\cB)\ar[r]^-{\delta}\ar[d]^-{\varepsilon}&
   \check{C}^{n+1}(\mf{U},\cB)\ar[r]^-{\delta}\ar[d]^-{\varepsilon}&\ldots\\
\ldots\ar[r]^-d&E^{n-1}\ar[r]^-d&E^{n}\ar[r]^-d&E^{n+1}\ar[r]^-d&\ldots
} \label{eq:qi1}
\end{equation}
gives another quasi-isomorphism of dga's.

%\% Where did my ``$\xi$'' go?  

%By using a partition of unity, as in \cite{BT:}, one may show that
%\begin{equation}
%\xymatrix@1{
%  0\ar[r]&\Lambda^{0,q}\ar[r]^-\xi&
% \check{C}^0(\mf{U},\cA^{0,q})\ar[r]^-\delta&
% \check{C}^1(\mf{U},\cA^{0,q})\ar[r]^-\delta&
% \check{C}^2(\mf{U},\cA^{0,q})\ar[r]^-\delta&\ldots,
%}
%\end{equation}
%is exact, where 
%This leads to another quasi-isomorphism of dga's:

An immediate consequence of this construction is Dolbeault's theorem:
\begin{equation}
  H^{0,q}_{\bar\partial}(X,B) \cong \check{H}^q(X,\cB).
\end{equation}
We have done a little more than just prove this fact however. We have
also given maps that induce this isomorphism and described how the
natural product structures are also mapped.

We may treat sheaf cohomology in a similar way. We use a double
complex given by
\begin{equation}
  \tilde{E}_0^{p,q} = \check{C}^p(\mf{U},\cI^q),\quad d=\delta+(-1)^p i_q
\end{equation}
A quasi-isomorphism analogous to (\ref{eq:qi1}) again follows. Since the 
sheaves $\cI^q$ are ``flabby'', one may also show that \cite{Hartshorne:}
\begin{equation}
\xymatrix@1{
0\ar[r]&\Gamma(\cI^q)\ar[r]^-\rho&\check{C}^0(\mf{U},\cI^q)\ar[r]^-\delta&
\check{C}^1(\mf{U},\cI^q)\ar[r]^-\delta&
\check{C}^2(\mf{U},\cI^q)\ar[r]^-\delta&\ldots
}
\end{equation}
is exact. This gives rise to yet another quasi-isomorphism of dga's:
\begin{equation}
\xymatrix@C=15mm{
\ldots\ar[r]^-{\Gamma(i_{n-2})}&\Gamma(\cI^{n-1})
\ar[r]^-{\Gamma(i_{n-1})}\ar[d]^-{\rho}
&\Gamma(\cI^{n})\ar[r]^-{\Gamma(i_{n})}\ar[d]^-{\rho}&
   \Gamma(\cI^{n+1})\ar[r]^-{\Gamma(i_{n+1})}\ar[d]^-{\rho}&\ldots\\
\ldots\ar[r]^-d&\tilde{E}^{n-1}\ar[r]^-d&\tilde{E}^{n}\ar[r]^-d&\tilde{E}^{n+1}
\ar[r]^-d&\ldots
} \label{eq:qi3}
\end{equation}
where 
\begin{equation}
  \tilde{E}^n = \bigoplus_{p+q=n} \tilde{E}_0^{p,q},
\end{equation}

Finally, note that the injective resolution (\ref{eq:Ires}) induces
a quasi-isomorphism from (\ref{eq:cechb}) to the bottom complex of
(\ref{eq:qi3}).  Thus all of the complexes we have discussed are
quasi-isomorphic to each other.  By Lemma~\ref{lem:ainfiso}, all
of the $\Ainf$-algebras we obtain are $\Ainf$-isomorphic to each other.
In particular, combining this result with the discussion at the end of
Section~\ref{s:sup}, we conclude that the superpotential of holomorphic
Chern-Simons theory is independent of the metric up to field redefinitions,
an expected property of the B-model.

\def\ddd{\mathsf{d}}
One is therefore free to recast the formulation of the topological
B-model into either \v Cech cohomology or sheaf cohomology. The idea
that one may use sheaf cohomology leads inexorably to the appearance
of the derived category $\DC(X)$, as reviewed in \cite{me:TASI-D}. So far we
have restricted attention to a single D-brane that fills $X$. One
extends this notion to any number of more general D-branes. The result is
that a D-brane is a complex of coherent sheaves
\begin{equation}
\cE^\bullet = \left(\xymatrix@1@C=15mm{
\ldots\ar[r]^-{\ddd_{n-2}}&\cE^{n-1}\ar[r]^-{\ddd_{n-1}}&
    \cE^{n}\ar[r]^-{\ddd_n}&\cE^{n+1}
    \ar[r]^-{d_{n+1}}&\ldots}\right).
\end{equation}
For the analysis of open strings between $\cE^\bullet$ and
$\cF^\bullet$ one replaces $\cB$ in the above discussion by the sheaf
$\sHom(\cE^m,\cF^n)$. One needs to extend the notation to cope with
these new complexes but this is an exercise only in bookkeeping and we
will spare the reader of this. The Hilbert space of open strings from
$\cE^\bullet$ to $\cF^\bullet$ is then given by ``hyperext'' groups:
\begin{equation}
  \bigoplus_n \Ext^n(\cE^\bullet,\cF^\bullet).
\end{equation}

Let us review exactly how to compute the $\Ainf$ structure of the
morphisms between objects in $\DC(X)$. Each object in $\DC(X)$ is
quasi-isomorphic to a complex of injective sheaves. We may view this
as an injective resolution of these objects. Without loss of
generality therefore, we may assume that the D-branes are given as a
complex of injective sheaves. Suppose, first, for simplicity, that we
have only one D-brane $\cE^\bullet$. The first row of (\ref{eq:qi3})
is then given by the complex with entries
\begin{equation}
   \bigoplus_p \Hom(\cE^p,\cE^{p+n}).
\end{equation}
If we denote an element of this group by $\sum_p f_{n,p}$, where
$f_{n,p}:\cE^p\to\cE^{p+n}$, then the differential for this complex is
given by
\begin{equation}
  \mf{d}_n f_{n,p} = \ddd_{p+n}\circ f_{n,p} - 
    (-1)^n f_{p+1,n}\circ \ddd_{p}.  \label{eq:mfd}
\end{equation}

For several D-branes, we write $\cE^\bullet = \cE_1^\bullet\oplus
\cE_2^\bullet\oplus\ldots$. The spaces of $\Hom$'s then break up into
direct sums and we may relabel everything in terms of morphisms
between the different objects $\cE_1^\bullet$, $\cE_2^\bullet$, etc.

This complex, together with the obvious product structure given by
composition, gives a dga. The cohomology of this complex gives the
Hilbert spaces of the various open string states. The method of
section \ref{s:Ainf} may then be used to compute the higher products
of the resulting $\Ainf$ category and thus we find the information
required for the superpotential.

%%%%%%%%%%%%%%%%%%%%%%%%%%%%%%%%%%%%%%%%%%%%%%%%%%%%%%%%%%%%%%%%%%%

\section{A Practical Method}  \label{s:prac}

\def\fd#1{\mathsf{#1}}

In the last section we achieved our primary goal. We rephrased the
question of how to compute the superpotential into a purely algebraic
one. There is no need to know the metric on $X$. Having said that, the
answer we obtained cannot really be viewed as a practical method of
computing the higher products. This is because it required finding an
injective resolution for each sheaf involved. While the existence of
injective resolutions is guaranteed (see \cite{Hartshorne:} for
example), an explicit construction is not usually forthcoming.

Instead we should use \v Cech cohomology as follows. In general, there
is a spectral sequence given by\footnote{This ``local to global''
  spectral sequence can be viewed as the Grothendieck spectral
  sequence (see \cite{Wei:hom} for example) applied to the composition
  of functors $\Gamma$ and $\sHom(\cE,-)$.}
\begin{equation}
  E_2^{p,q} = H^p(X,\sExt^q(\cE,\cF)),
\end{equation}
that converges to $\Ext^{p+q}(\cE,\cF)$. If $\cE$ is locally-free,
then $\sExt^q(\cE,\cF)=0$ for $q>0$ and therefore
\begin{equation}
\begin{split}
  \Ext^n(\cE,\cF) &= H^n(X,\sHom(\cE,\cF))\\
  &= \check{H}^n(X,\sHom(\cE,\cF)).
\end{split}
\end{equation}
Now any coherent sheaf on a smooth $X$, has a locally-free resolution
and so we are free to represent any object of $\DC(X)$ by a complex of
locally-free sheaves. Unlike the case of injectives representations,
it is usually straightforward to compute a locally-free representation
of a given object in $\DC(X)$.

So we proceed as follows. Suppose, again, for simplicity of notation,
that we have a single D-brane which is represented by a complex
$\cE^\bullet$ of locally-free sheaves. We have a complex with entries denoted:
\begin{equation}
  \sHom^q(\cE^\bullet,\cE^\bullet) =
  \bigoplus_m\sHom(\cE^m,\cE^{m+q}),
\end{equation}
and a differential $\mf{d}_q$ given by (\ref{eq:mfd}). 
Now build a double complex with entries
\begin{equation}
  \bigoplus_{p+q=n} 
    \check{C}^p\left(\mf{U},\sHom^q(\cE^\bullet,\cE^\bullet)\right),
\end{equation}
of degree $n$, and differential $d=\delta+(-1)^{p}\mf{d}_{q}$, where
$\mf{d}_q$ is given by (\ref{eq:mfd}). 

There is a natural product, given by
the \v Cech cup product combined with composition of maps in the
$\sHom$ sheaves. Suppose
\begin{equation}
\begin{split}
  \fd a &\in
  \check{C}^p\left(\mf{U},\sHom^q(\cE^\bullet,\cE^\bullet)\right)\\
  \fd b &\in
  \check{C}^r\left(\mf{U},\sHom^s(\cE^\bullet,\cE^\bullet)\right)
\end{split}
\end{equation}
and let us denote the natural composition $\fd a \cdot \fd b$. This
composition fails to satisfy the required Leibniz rule and instead we
define a product
\begin{equation}
  \fd a \star \fd b = (-1)^{qr} \fd a\cdot \fd b.
\end{equation}
This new product gives us the structure of a dga. 

By the same methods that were employed above, this dga is again
quasi-isomorphic to all those considered in section \ref{s:CS}.  The
presentation of the dga is actually perfectly practical to use, at
least in relatively simple cases as we now demonstrate.

In order to compute \v Cech cohomology, we need an open cover of $X$
that is sufficiently fine. That is, we need all the open sets, and all
the intersections of the open sets, to have trivial sheaf
cohomology. A sufficient condition for this is that the open sets and
their intersections be {\em affine\/} \cite{Hartshorne:}. A space is
affine if it can be written as the solution of a set of algebraic
equations in $\C^n$, for some $n$.

For example, consider the projective space $\P^n$ with homogeneous
coordinates $[z_0,z_1,\ldots,\allowbreak z_n]$. The usual patches
$U_i$, isomorphic to
$\C^n$, are defined by $z_i\neq 0$. Let
\begin{equation}
  U_{i_0i_1\ldots i_p} = U_{i_0}\cap U_{i_1} \cap\ldots\cap U_{i_p}.
\end{equation}
The space $U_{i_0,i_1\ldots i_p}\cong (C^*)^p\times\C^{n-p}$ is 
isomorphic to the affine variety
defined by $z_{i_0}z_{i_1}\ldots z_{i_p}=1$ in $\C^{n+1}$. Thus this
cover is good enough for our purposes. Note that any algebraic variety
defined within this $\P^n$ can also use this cover.

Before giving some examples, let us fix notation for \v Cech
cochains. In our examples, our sheaves $\cF$ will be vector bundles
which have been trivialized over each $U_i$, so that sections of $\cF$ over
$U_i$ can and will be identified with tuples of functions on $U_i$.
As usual, when we change trivializations, we must multiply by an 
appropriate transition function.

For the higher cochains we will make a notational choice in describing
elements of $\cF(U_{i_0,i_1,\ldots, i_p})$ since many different
trivializations are possible in general.  Our choice will consistently
be to choose the trivialization over $U_{i_0}$.  So
$(f)_{i_0,i_1,\ldots,i_p}$ denotes a section of
$\cF(U_{i_0,i_1,\ldots, i_p})$ over $U_{i_0,i_1,\ldots, i_p}$,
expressed as a vector of functions using the given trivialization of
$\cF$ over $U_{i_0}$.

As a special case, if a 0-chain is a
global section, i.e., a 0-cocycle, then we denote it simply by $f$,
when $f$ denotes its expression in the $U_0$ trivialization.

For example, for the sheaf $\O(n)$ on $\P^1$, $(f)_{01}$ is a 1-cochain
given by $f$ in terms of variables for $U_0$. It will therefore be
given by $(f)_{01}.(z_0/z_1)^n$ in terms of variables in the patch $U_1$.

\subsection{The conifold point of type $(-1,-1)$}

For the first example consider a 3-brane (i.e., a point-like object in
the compact directions) on a conifold point obtained by contracting a
curve $C\cong\P^1$ with normal bundle $\O_C(-1)\oplus\O_C(-1)$. As
explained in \cite{me:point}, this 3-brane is marginally stable with
respect to decay into $\O_C$ and $\O_C(-1)[1]$. Thus, if we considered
$N$ coincident 3-branes at this conifold point, we would have a
$\GU(N)\times\GU(N)$ quiver gauge theory.

The superpotential for this case is known. It is computed in
\cite{KW:coni,MP:AdS} by somewhat indirect means. This will provide a
useful check for our method of computation. One may also regard our
computation as a more rigorous proof of the result.

To produce a local model for this case, let $X$ be the total space of
the normal bundle $\O_C(-1)\oplus\O_C(-1)$. Thus we have bundle map
$\pi:X \to C$. An affine open cover of $X$ is then given by two
patches: $U_0$, with coordinates $(x,y_1,y_2)$; and $U_1$, with
coordinates $(w,z_1,z_2)$. The transition functions are obviously
\begin{equation}
\begin{split}
  w &= x^{-1}\\
  z_1 &= xy_1\\
  z_2 &= xy_2
\end{split}
\end{equation}
Now $\O_C$ is not a locally-free sheaf on $X$. Define
$\O(1)=\pi^*\O_C(1)$. We then have an exact sequence
\begin{equation}
\xymatrix@1@C=15mm{
  0\ar[r]&\O(2)\ar[r]
     \ar[r]^-{\left(\begin{smallmatrix}-y_2\\y_1\end{smallmatrix}\right)}_-{
     \left(\begin{smallmatrix}-z_2\\z_1\end{smallmatrix}\right)}&
  \O(1)\oplus\O(1)
     \ar[r]^-{\left(\begin{smallmatrix}y_1&y_2\end{smallmatrix}\right)}_-{
     \left(\begin{smallmatrix}z_1&z_2\end{smallmatrix}\right)}&
  \O\ar[r]&\O_C\ar[r]&0,
}
\end{equation}
where we have given the explicit sheaf maps in both patches. This
provides the locally-free resolution of $\O_C$, and thus $\O_C(-1)[1]$
too by tensoring the resolution by $\O(-1)$ and shifting one place to the
left.

$\Ext^1(\O_C(-1)[1],\O_C)$ and $\Ext^1(\O_C,\O_C(-1)[1])$ are both
isomorphic to $\C^2$. Thus we have a quiver:
\begin{equation}
\begin{xy} <1.0mm,0mm>:
  (0,0)*{\circ}="a",(40,0)*{\circ}="b",
  (-12,0)*{\O_C(-1)[1]},(45,0)*{\O_C},
  \ar@{->}@/^8mm/|{\fd a} "a";"b"
  \ar@{->}@/^3mm/|{\fd b} "a";"b"
  \ar@{->}@/^8mm/|{\fd d} "b";"a"
  \ar@{->}@/^3mm/|{\fd c} "b";"a"
\end{xy}
\end{equation}

Open strings correspond to maps which are $d$-closed.  In turns out
that in this example we may represent all the required $d$-closed maps
by maps which are both $\delta$-closed and $\mf{d}$-closed as we now
see explicitly. The classes in $\Ext^1(\O_C(-1)[1],\O_C)$ are
represented by elements of $\check{C}^0(\mf{U},\sHom^1(\O_C(-1)[1],\O_C))$ as
follows. Using the notation described above, let one
generator of this group, denoted $\fd a$, be represented by
\begin{equation}
\xymatrix{
\O(1)\ar[r]^-{\left(\begin{smallmatrix}-y_2\\y_1\end{smallmatrix}\right)}
\ar[d]^1&
\O\oplus\O\ar[r]^-{\left(\begin{smallmatrix}y_1&y_2\end{smallmatrix}\right)}
\ar[d]^-{-\left(\begin{smallmatrix}1&0\\0&1\end{smallmatrix}\right)}&
\O(-1)\ar[d]^1\\
\O(2)\ar[r]^-{\left(\begin{smallmatrix}-y_2\\y_1\end{smallmatrix}\right)}&
\O(1)\oplus\O(1)
\ar[r]^-{\left(\begin{smallmatrix}y_1&y_2\end{smallmatrix}\right)}&
\O
}
\end{equation}
and $\fd b$ by
\begin{equation}
\xymatrix{
\O(1)\ar[r]^-{\left(\begin{smallmatrix}-y_2\\y_1\end{smallmatrix}\right)}
\ar[d]^x&
\O\oplus\O\ar[r]^-{\left(\begin{smallmatrix}y_1&y_2\end{smallmatrix}\right)}
\ar[d]^-{-\left(\begin{smallmatrix}x&0\\0&x\end{smallmatrix}\right)}&
\O(-1)\ar[d]^x\\
\O(2)\ar[r]^-{\left(\begin{smallmatrix}-y_2\\y_1\end{smallmatrix}\right)}&
\O(1)\oplus\O(1)
\ar[r]^-{\left(\begin{smallmatrix}y_1&y_2\end{smallmatrix}\right)}&
\O.
}
\end{equation}
\vspace{1cm}

Next, the two generators of $\Ext^1(\O_C,\O_C(-1)[1])$ can be
represented by elements of
$\check{C}^1(\mf{U},\sHom^0(\O_C,\O_C(-1)[1])$. Let $\fd c$ be
represented by
\begin{equation}
\label{eq:crep}
\xymatrix@R+15mm{
&\O(2)\ar[r]^-{\left(\begin{smallmatrix}-y_2\\y_1\end{smallmatrix}\right)}
\ar[d]^{\left(\begin{smallmatrix}0\\-\frac1x\end{smallmatrix}\right)_{01}}&
\O(1)\oplus\O(1)
\ar[r]^-{\left(\begin{smallmatrix}y_1&y_2\end{smallmatrix}\right)}
\ar[d]^{\left(\begin{smallmatrix}\frac1x&0\end{smallmatrix}\right)_{01}}&
\O\\
\O(1)\ar[r]^-{\left(\begin{smallmatrix}-y_2\\y_1\end{smallmatrix}\right)}&
\O\oplus\O
\ar[r]^-{\left(\begin{smallmatrix}y_1&y_2\end{smallmatrix}\right)}&
\O(-1)
}
\end{equation}
and $\fd d$ by
\begin{equation}
\label{eq:drep}
\xymatrix@R+15mm{
&\O(2)\ar[r]^-{\left(\begin{smallmatrix}-y_2\\y_1\end{smallmatrix}\right)}
\ar[d]^{\left(\begin{smallmatrix}\frac1x\\0\end{smallmatrix}\right)_{01}}&
\O(1)\oplus\O(1)
\ar[r]^-{\left(\begin{smallmatrix}y_1&y_2\end{smallmatrix}\right)}
\ar[d]^{\left(\begin{smallmatrix}0&\frac1x\end{smallmatrix}\right)_{01}}&
\O\\
\O(1)\ar[r]^-{\left(\begin{smallmatrix}-y_2\\y_1\end{smallmatrix}\right)}&
\O\oplus\O
\ar[r]^-{\left(\begin{smallmatrix}y_1&y_2\end{smallmatrix}\right)}&
\O(-1)
}
\end{equation}

Finally, the generator of $\Ext^3(\O_C(-1)[1],\O_C(-1)[1])$ can be
represented by a 
1-cochain in $\check{C}^1(\mf{U},\sHom^2(\O_C(-1)[1],\O_C(-1)[1]))$:
\begin{equation}
\xymatrix@R+15mm{
&&\O(1)\ar[r]^-{\left(\begin{smallmatrix}-y_2\\y_1\end{smallmatrix}\right)}
\ar[d]^{(\frac1x)_{01}}&
\O\oplus\O
\ar[r]^-{\left(\begin{smallmatrix}y_1&y_2\end{smallmatrix}\right)}&
\O(-1)\\
\O(1)\ar[r]^-{\left(\begin{smallmatrix}-y_2\\y_1\end{smallmatrix}\right)}&
\O\oplus\O
\ar[r]^-{\left(\begin{smallmatrix}y_1&y_2\end{smallmatrix}\right)}&
\O(-1)
}  \label{eq:Ext3}
\end{equation}

The composition $\fd c\star \fd a$ gives a map
\begin{equation}
\xymatrix@R+15mm{
&\O(1)\ar[r]^-{\left(\begin{smallmatrix}-y_2\\y_1\end{smallmatrix}\right)}
\ar[d]^{\left(\begin{smallmatrix}0\\-\frac1x\end{smallmatrix}\right)_{01}}&
\O\oplus\O
\ar[r]^-{\left(\begin{smallmatrix}y_1&y_2\end{smallmatrix}\right)}
\ar[d]^{\left(\begin{smallmatrix}-\frac1x&0\end{smallmatrix}\right)_{01}}&
\O(-1)\\
\O(1)\ar[r]^-{\left(\begin{smallmatrix}-y_2\\y_1\end{smallmatrix}\right)}&
\O\oplus\O
\ar[r]^-{\left(\begin{smallmatrix}y_1&y_2\end{smallmatrix}\right)}&
\O(-1)
}
\end{equation}

This is exact. To be precise, $\fd c\star \fd a$ is a \v Cech
coboundary of the map which is zero in patch 0 and in patch 1 given by
the chain map
\begin{equation}
\xymatrix@R+15mm{
&\O(1)\ar[r]^-{\left(\begin{smallmatrix}-y_2\\y_1\end{smallmatrix}\right)}
\ar[d]^{\left(\begin{smallmatrix}0\\-1\end{smallmatrix}\right)_{1}}&
\O\oplus\O
\ar[r]^-{\left(\begin{smallmatrix}y_1&y_2\end{smallmatrix}\right)}
\ar[d]^{\left(\begin{smallmatrix}-1&0\end{smallmatrix}\right)_{1}}&
\O(-1)\\
\O(1)\ar[r]^-{\left(\begin{smallmatrix}-y_2\\y_1\end{smallmatrix}\right)}&
\O\oplus\O
\ar[r]^-{\left(\begin{smallmatrix}y_1&y_2\end{smallmatrix}\right)}&
\O(-1) 
} \label{eq:CA}
\end{equation}
Thus, from (\ref{eq:m2}), $m_2(\fd c,\fd a)=0$ and $f_2(\fd c,\fd a)$
is given by minus (\ref{eq:CA}).

Now compose this with $\fd b$ to form $\fd b\star f_2(\fd c,\fd a)$
given by the \v Cech 0-chain
\begin{equation}
\xymatrix@R+15mm{
&\O(1)\ar[r]^-{\left(\begin{smallmatrix}-y_2\\y_1\end{smallmatrix}\right)}
\ar[d]^{\left(\begin{smallmatrix}0\\-1\end{smallmatrix}\right)_{1}}&
\O\oplus\O
\ar[r]^-{\left(\begin{smallmatrix}y_1&y_2\end{smallmatrix}\right)}
\ar[d]^{\left(\begin{smallmatrix}1&0\end{smallmatrix}\right)_{1}}&
\O(-1)\\
\O(2)\ar[r]^-{\left(\begin{smallmatrix}-y_2\\y_1\end{smallmatrix}\right)}&
\O(1)\oplus\O(1)
\ar[r]^-{\left(\begin{smallmatrix}y_1&y_2\end{smallmatrix}\right)}&
\O
}
\end{equation}
This corresponds to one of the terms needed to compute $m_3(\fd b,\fd
c,\fd a)$.

Similarly, a computation for $f_2(\fd b, \fd c)\star \fd a$ yields
\begin{equation}
\xymatrix@R+15mm{
&\O(1)\ar[r]^-{\left(\begin{smallmatrix}-y_2\\y_1\end{smallmatrix}\right)}
\ar[d]^{\left(\begin{smallmatrix}0\\-1\end{smallmatrix}\right)_{0}}&
\O\oplus\O
\ar[r]^-{\left(\begin{smallmatrix}y_1&y_2\end{smallmatrix}\right)}
\ar[d]^{\left(\begin{smallmatrix}1&0\end{smallmatrix}\right)_{0}}&
\O(-1)\\
\O(2)\ar[r]^-{\left(\begin{smallmatrix}-y_2\\y_1\end{smallmatrix}\right)}&
\O(1)\oplus\O(1)
\ar[r]^-{\left(\begin{smallmatrix}y_1&y_2\end{smallmatrix}\right)}&
\O
}
\end{equation}

Remembering the rule (\ref{eq:sign}), from (\ref{eq:m3}) we see that
$m_3(\fd b,\fd c,\fd a)$ is equal to $-\fd b\star f_2(\fd c,\fd
a)-f_2(\fd b, \fd c)\star \fd a$ and is thus given by the following
globally defined map which represents a generator of
$\Ext^2(\O_C(-1)[1],\O_C)$:
\begin{equation}
\xymatrix@R+15mm{
&\O(1)\ar[r]^-{\left(\begin{smallmatrix}-y_2\\y_1\end{smallmatrix}\right)}
\ar[d]^{\left(\begin{smallmatrix}0\\1\end{smallmatrix}\right)}&
\O\oplus\O
\ar[r]^-{\left(\begin{smallmatrix}y_1&y_2\end{smallmatrix}\right)}
\ar[d]^{\left(\begin{smallmatrix}-1&0\end{smallmatrix}\right)}&
\O(-1)\\
\O(2)\ar[r]^-{\left(\begin{smallmatrix}-y_2\\y_1\end{smallmatrix}\right)}&
\O(1)\oplus\O(1)
\ar[r]^-{\left(\begin{smallmatrix}y_1&y_2\end{smallmatrix}\right)}&
\O
}
\end{equation}
When composed with $\fd d$ this gives the $\Ext^3$ of (\ref{eq:Ext3})
but when composed with $\fd c$ it gives zero. Thus $m_3(\fd b,\fd
c,\fd a)$ is Serre dual to $\fd d$.  Denoting by $A$ the $N=1$
superfield dual to $\fd a$ etc., we thus have a term in the
superpotential equal to $\Tr(BCAD)$.

Composing the other way to find $m_3(\fd a,\fd c,\fd b)$ gives a
similar result except for a sign. There are no other higher products
and so the total is, in agreement with \cite{KW:coni}:
$$
W=\Tr(BCAD-ACBD).
$$
One is also free to do nonlinear field redefinitions as discussed at
the end of section \ref{s:sup}.

\subsection{A $\P^1$ with higher obstructions.}

Our next example is a conifold-like point associated with an obstructed
$\P^1$ with normal bundle $\O\oplus\O(-2)$.  An example of such a
$\P^1$ can be given explicitly in patches using the transition
functions

\begin{equation}
\begin{split}
w &= x^{-1}\\
z_1 &= x^2y_1 + xy_2^n\\
z_2 &= y_2
\end{split}
\end{equation}
with $n\ge2$ (the $n=1$ case can be identified with the resolved conifold
after a change of variables).

The quiver for a decay of a 3-brane into $\O_C$ and $\O_C(-1)[1]$ in
this case is given by
\begin{equation}
\begin{xy} <1.0mm,0mm>:
  (0,0)*{\circ}="a",(40,0)*{\circ}="b",
  (-8,-10)*{\O_C(-1)[1]},(40,-5)*{\O_C},
  \ar@{->}@/^8mm/|{\fd a} "a";"b"
  \ar@{->}@/^3mm/|{\fd b} "a";"b"
  \ar@{->}@/^8mm/|{\fd d} "b";"a"
  \ar@{->}@/^3mm/|{\fd c} "b";"a"
  \ar@{->}@`{(45,-10),(50,-3),(50,3),(45,10)}|{\fd x}  "b";"b"
  \ar@{->}@`{(-5,-10),(-10,-3),(-10,3),(-5,10)}|{\fd y}  "a";"a"
\end{xy} \label{eq:q2}
\end{equation}

A locally-free resolution of $\O_C$ is given by
\begin{equation}
\xymatrix@1@C=25mm{
\O
\ar[r]^{\left(\begin{smallmatrix}y_2\\-1\\x\end{smallmatrix}\right)}
&{\begin{matrix}\O\\\oplus\\\O(1)\\\oplus\\\O(1)\end{matrix}}
\ar[r]^{\left(\begin{smallmatrix}1&y_2&0\\-x&0&y_2\\
   -y_2^{n-1}&-s&-y_1\end{smallmatrix}\right)}
&{\begin{matrix}\O(1)\\\oplus\\\O(1)\\\oplus\\\O\end{matrix}}
\ar[r]^{\left(\begin{smallmatrix}s&y_1&y_2\end{smallmatrix}\right)}
&\O\ar[r]&\O_C. 
}\label{eq:freeX}
\end{equation}
where $s=xy_1+y_2^n$.  

In constructing this resolution, the bundles
$\O(n)$ which appear were chosen so that all maps appearing in the
resolution remain holomorphic in the $U_1$ after changing coordinates
and multiplying by the transition function $x^{-n}$ of $\O(n)$.  For
example, $s$ is given as a section of $\O(-1)$.  In $U_1$ coordinates,
this becomes $z_1=xs$ which is holomorphic.

Note that the sections $s,y_1,y_2$ have been chosen to generate the ideal
of all functions vanishing on $C$ in both patches.  In $U_0$, the sections
$y_1$ and $y_2$ already suffice to generate the ideal.  In $U_1$, these
sections become $z_1,wz_1-z_2^n,z_2$ respectively, and now $z_1$ and $z_2$
already suffice.  Note in particular that it was necessary to include the
section $s$, as $y_1,y_2$ would not have sufficed:
in $U_1$ these get identified with $wz_1-z_2^n,z_2$ which fail to 
generate the ideal of the curve at $(w,z_1,z_2)=(0,0,0)$.

Define $\fd x$ to be the
following generator of $\Ext^1(\O_C,\O_C)\cong\C$:
\begin{equation}
\xymatrix@C+20mm@R+20mm{
&\O
\ar[d]^{\left(\begin{smallmatrix}1\\0\\0\end{smallmatrix}\right)}
\ar[r]^{\left(\begin{smallmatrix}y_2\\-1\\x\end{smallmatrix}\right)}
&{\begin{matrix}\O\\\oplus\\\O(1)\\\oplus\\\O(1)\end{matrix}}
\ar[d]|{\left(\begin{smallmatrix}0&1~&0\\0&0~&1\\
   y_2^{n-2}&0~&0\end{smallmatrix}\right)}
\ar[r]^{\left(\begin{smallmatrix}1&y_2&0\\-x&0&y_2\\
   -y_2^{n-1}&-s&-y_1\end{smallmatrix}\right)}
&{\begin{matrix}\O(1)\\\oplus\\\O(1)\\\oplus\\\O\end{matrix}}
\ar[d]^{\left(\begin{smallmatrix}0&0&1\end{smallmatrix}\right)}
\ar[r]^{\left(\begin{smallmatrix}s&y_1&y_2\end{smallmatrix}\right)}
&\O
\\
\O
\ar[r]^{\left(\begin{smallmatrix}y_2\\-1\\x\end{smallmatrix}\right)}
&{\begin{matrix}\O\\\oplus\\\O(1)\\\oplus\\\O(1)\end{matrix}}
\ar[r]^{\left(\begin{smallmatrix}1&y_2&0\\-x&0&y_2\\
   -y_2^{n-1}&-s&-y_1\end{smallmatrix}\right)}
&{\begin{matrix}\O(1)\\\oplus\\\O(1)\\\oplus\\\O\end{matrix}}
\ar[r]^{\left(\begin{smallmatrix}s&y_1&y_2\end{smallmatrix}\right)}
&\O
} 
\end{equation}

From now on, for brevity, let us refer to the sheaves in the locally-free
resolution (\ref{eq:freeX}) as $\cF_i$.  
$\Ext^3(\O_C,\O_C)$ is represented by the 0-cochain:
\begin{equation}
\xymatrix{
&&&\cF_3\ar[r]\ar[d]^1&\cF_2\ar[r]&\cF_1\ar[r]&\cF_0\\
\cF_3\ar[r]&\cF_2\ar[r]&\cF_1\ar[r]&\cF_0
}  \label{eq:Ext3a}
\end{equation}
or, equivalently, by the 1-cochain:
\begin{equation}
\xymatrix@R+10mm@C+0mm{
&&\cF_3\ar[r]
\ar[d]^0&
\cF_2\ar[r]
\ar[d]^{\left(\begin{smallmatrix}0&1&\frac1x
   \end{smallmatrix}\right)_{01}}&
\cF_1\ar[r]&
\cF_0\\
\cF_3\ar[r]&\cF_2\ar[r]&\cF_1\ar[r]&\cF_0
}
\end{equation}
These two choices differ by a $d$-boundary. We will use the representative
(\ref{eq:Ext3a}) to describe the $\Ainf$-algebra via Kadeishvili's theorem.

We compute $\fd x\star \fd x$ to be $\fd J_{n-2}$, where $\fd J_p\in
\Ext^2(\O_C,\O_C)$ is defined as
\begin{equation}
\xymatrix@C+8mm@R+6mm{
&&\cF_3
\ar[d]^{\left(\begin{smallmatrix}0\\0\\y_2^p\end{smallmatrix}\right)}
\ar[r]
&\cF_2
\ar[d]^{\left(\begin{smallmatrix}y_2^p&0&0\end{smallmatrix}\right)}\ar[r]
&\cF_1\ar[r]
&\cF_0
\\
\cF_3\ar[r]&\cF_2\ar[r]&\cF_1\ar[r]&\cF_0
} \label{eq:X2}
\end{equation}
But, if $p\geq1$ then 
\begin{equation}
  \fd J_p = d\fd K_{p-1}, \label{eq:J=dK}
\end{equation}
where $\fd K_p$ is given by
\begin{equation}
\xymatrix@C+8mm@R+6mm{
&\cF_3
\ar[d]^{\left(\begin{smallmatrix}0\\0\\0\end{smallmatrix}\right)}\ar[r]
&\cF_2
\ar[d]|{\left(\begin{smallmatrix}0&0~&0\\0&0~&0\\
   y_2^p&0~&0\end{smallmatrix}\right)}\ar[r]
&\cF_1
\ar[d]^{\left(\begin{smallmatrix}0&0&0\end{smallmatrix}\right)}
\ar[r]
&\cF_0
\\
\cF_3\ar[r]&\cF_2\ar[r]&\cF_1\ar[r]&\cF_0
} 
\end{equation}
It is now easy to see that
\begin{equation}
 \fd J_p = \fd x\star \fd K_p + \fd K_p\star \fd x. \label{eq:JK}
\end{equation}
Applying (\ref{eq:Amorph}) and using the fact that
$\fd K_i\star\fd K_j=0$, it follows that we can choose
\begin{equation}
\left.\begin{aligned}
  f_k(\fd x,\fd x,\ldots,\fd x) &= (-1)^{\frac{k(k-1)}2}\fd K_{n-k-1} \\
  m_k(\fd x,\fd x,\ldots,\fd x) &= 0 \end{aligned}
  \right\}\quad\hbox{for $2\le k<n$.}
\end{equation}
and
\begin{equation}
  m_n(\fd x,\fd x,\ldots,\fd x) = -(-1)^{\frac{n(n-1)}2}\fd J_0.
\end{equation}
But $\fd J_0$ composed with $\fd x$ is the generator of $\Ext^3$ given
in (\ref{eq:Ext3a}) so we have a term in the superpotential equal to
$-(-1)^{\frac{n(n-1)}2}X^{n+1}$. Similarly we obtain a contribution
$-(-1)^{\frac{n(n-1)}2}Y^{n+1}$ to the superpotential.

The next few arrows in (\ref{eq:q2}) are given by:
\begin{equation}
\begin{xy} <1em,0em>:
(0,0)*\xybox{\xymatrix{
  \cF_3(-1)\ar[r]\ar[d]^{-{1}}&\cF_2(-1)
  \ar[r]\ar[d]^{\mathbf{1}}&\cF_1(-1)\ar[r]\ar[d]^{-\mathbf{1}}
  &\cF_0(-1)\ar[d]^1\\
  \cF_3\ar[r]&\cF_2\ar[r]&\cF_1\ar[r]&\cF_0}},
(-4,-1.8)*{\fd a =}
\end{xy}
\end{equation}
\begin{equation}
\begin{xy} <1em,0em>:
(0,0)*\xybox{\xymatrix{
  \cF_3(-1)\ar[r]\ar[d]^{-{x}}&\cF_2(-1)
  \ar[r]\ar[d]^{x.\mathbf{1}}&\cF_1(-1)\ar[r]\ar[d]^{-x.\mathbf{1}}
  &\cF_0(-1)\ar[d]^x\\
  \cF_3\ar[r]&\cF_2\ar[r]&\cF_1\ar[r]&\cF_0}},
(-4,-1.8)*{\fd b =}
\end{xy}
\end{equation}
\begin{equation}
\begin{xy} <1em,0em>:
(0,0)*\xybox{\xymatrix@R+10mm@C+0mm{
&\cF_3\ar[r]\ar[d]^0&\cF_2\ar[r]
\ar[d]^{\left(\begin{smallmatrix}0&0&0\\0&0&0\\0&1&\frac1x
   \end{smallmatrix}\right)_{01}}&
\cF_1\ar[r]
\ar[d]^{\left(\begin{smallmatrix}1&\frac1x&0\end{smallmatrix}\right)_{01}}&
\cF_0\\
\cF_3(-1)\ar[r]&\cF_2(-1)\ar[r]&\cF_1(-1)\ar[r]&\cF_0(-1)}},
(-4,-1.8)*{\fd c =}
\end{xy}
\end{equation}

A new feature appears when we try to write down the final map $\fd
d$. Unlike the above cases we cannot use a single map with $\delta\fd
d= \mf{d}\fd{d}=0$. Instead we need to write $\fd d$ as a sum $f+h$,
where $f$ is a class in $\check{C}^1(\mf{U},\sHom^0(\O,\O(-1)[1]))$:
\begin{equation}
\xymatrix@R+10mm@C+0mm{
&\cF_3\ar[r]\ar[d]^0&\cF_2\ar[r]
\ar[d]^{\left(\begin{smallmatrix}0&0&0\\0&0&0\\0&-\frac1x&0
   \end{smallmatrix}\right)_{01}}&
\cF_1\ar[r]
\ar[d]^{\left(\begin{smallmatrix}-\frac1x&0&0\end{smallmatrix}\right)_{01}}&
\cF_0\\
\cF_3(-1)\ar[r]&\cF_2(-1)\ar[r]&\cF_1(-1)\ar[r]&\cF_0(-1)
}
\end{equation}
and $h$ is a class in $\check{C}^0(\mf{U},\sHom^1(\O,\O(-1)[1]))$:
\begin{equation}
\xymatrix@R+10mm@C+0mm{
&&\cF_3\ar[r]
\ar[d]^{\left(\begin{smallmatrix}0\\0\\-1\end{smallmatrix}\right)_{1}}&
\cF_2\ar[r]
\ar[d]^{\left(\begin{smallmatrix}1&0&0
   \end{smallmatrix}\right)_{1}}&
\cF_1\ar[r]&
\cF_0\\
\cF_3(-1)\ar[r]&\cF_2(-1)\ar[r]&\cF_1(-1)\ar[r]&\cF_0(-1)
}
\end{equation}
Then $d\fd d=-\mf{d}f+\delta h=0$ as required.

A straight-forward computation, whose details we omit, then yields
\begin{equation}
  W = \Tr\Bigl(-(-1)^{\frac{n(n-1)}2}X^{n+1} 
  -(-1)^{\frac{n(n-1)}2}Y^{n+1} 
- XAC + XBD - YCA + YDB\Bigr)
\end{equation}
in agreement with \cite{CKV:quiv} for example.

\subsection{A new example of type $(1,-3)$}

Here we consider a 5-brane wrapping a $\P^1$ locally given by
\begin{equation}
\begin{split}
  w &= x^{-1}\\
  z_1 &= x^3y_1+y_2^2\\
  z_2 &= x^{-1}y_2
\end{split}\label{eq:trans3}
\end {equation}
This curve cannot be contracted and so we do not consider 3-brane
decay in this case. There are 2 massless open strings beginning and
ending on this $\P^1$ and the moduli space is again obstructed as we
see below.  It already follows from \cite{Katz:super} that the moduli
space can be defined as the critical point locus of a superpotential-like
function $XY^2$, but no claim was made there that this coincides with the
physical superpotential.  Our computations will show that this is indeed
the physical superpotential.

The equation $w^2z_1-z_2^2=xy_1$ shows that $y_1$ can be identified
with a global section of $\O(-1)$.  Similarly, the last equation in
(\ref{eq:trans3}) shows that $y_2$ can be identified with a section of
$\O(1)$.

A resolution of $\O_C$ yields the following complex of
locally-free sheaves representing the D-brane:
\begin{equation}
\xymatrix@1@C=25mm{
\O(-3)
\ar[r]^{\left(\begin{smallmatrix}y_2\\-x^3\\-1\end{smallmatrix}\right)}
&{\begin{matrix}\O(-2)\\\oplus\\\O\\\oplus\\\O(-1)\end{matrix}}
\ar[r]^{\left(\begin{smallmatrix}x^3&y_2&0\\y_2&-y_1&z_1\\
   -1&0&-y_2\end{smallmatrix}\right)}
&{\begin{matrix}\O(1)\\\oplus\\\O(-1)\\\oplus\\\O\end{matrix}}
\ar[r]^{\left(\begin{smallmatrix}y_1&y_2&z_1\end{smallmatrix}\right)}
&\O
%\ar[r]&\O_C. 
}.\label{eq:free3}
\end{equation}
In (\ref{eq:free3}), we have used $z_1$ as an abbreviation for its expression
$x^3y_1+y_2^2$ in the $U_0$ patch.

Define $\fd x$ and $\fd y$ to be the
following generators of $\Ext^1(\O_C,\O_C)\cong\C$:
\begin{equation}
\begin{xy} <1em,0em>:
(0,0)*\xybox{\xymatrix@C+8mm@R+6mm{
&\cF_3
\ar[d]^{\left(\begin{smallmatrix}1\\0\\0\end{smallmatrix}\right)}\ar[r]
\ar[r]
&\cF_2
\ar[d]^{\left(\begin{smallmatrix}0&1&0\\-1&0&0\\
   0&0&-1\end{smallmatrix}\right)}\ar[r]
&\cF_1
\ar[d]^{\left(\begin{smallmatrix}0&1&0\end{smallmatrix}\right)}
\ar[r]
&\cF_0
\\
\cF_3\ar[r]&\cF_2\ar[r]&\cF_1\ar[r]&\cF_0}},
(-4,-1.8)*{\fd x =}
\end{xy}
\end{equation}
\begin{equation}
\begin{xy} <1em,0em>:
(0,0)*\xybox{\xymatrix@C+8mm@R+6mm{
&\cF_3
\ar[d]^{\left(\begin{smallmatrix}x\\0\\0\end{smallmatrix}\right)}\ar[r]
&\cF_2
\ar[d]^{\left(\begin{smallmatrix}0&x&0\\-x&0&0\\
   0&0&-x\end{smallmatrix}\right)}\ar[r]
&\cF_1
\ar[d]^{\left(\begin{smallmatrix}0&x&0\end{smallmatrix}\right)}
\ar[r]
&\cF_0
\\
\cF_3\ar[r]&\cF_2\ar[r]&\cF_1\ar[r]&\cF_0}}, 
(-4,-1.8)*{\fd y =}
\end{xy}
\end{equation}

In the $U_0$ patch, we can simply write $\fd y=x\fd x$.
Note that the entries of the above matrices remain holomorphic in the $U_1$
patch, and prevent us from multiplying the entries by any higher powers of $x$.
We compute $\fd x \star\fd x$ to be
\begin{equation}
\xymatrix@C+8mm@R+6mm{
&&\cF_3\ar[r]\ar[d]^{\left(\begin{smallmatrix}0\\-1\\0\end{smallmatrix}
\right)}
&\cF_2
\ar[d]^{\left(\begin{smallmatrix}-1&0&0\end{smallmatrix}\right)}\ar[r]
&\cF_1\ar[r]
&\cF_0
\\
\cF_3\ar[r]&\cF_2\ar[r]&\cF_1\ar[r]&\cF_0
} 
\end{equation}
From this, we compute immediately that $\fd y\star\fd x=x(\fd x\star\fd x)$ and
$\fd y\star\fd y=x^2(\fd x\star\fd x)$.

We now note that $\fd x \star\fd x$ is exact, given by $\mf{d}$
applied to
\begin{equation}
\xymatrix@C+8mm@R+6mm{
&\cF_3
\ar[d]^{\left(\begin{smallmatrix}0\\0\\0\end{smallmatrix}\right)}\ar[r]
&\cF_2
\ar[d]^{\left(\begin{smallmatrix}0&0&0\\0&0&1\\
   0&0&0\end{smallmatrix}\right)}\ar[r]
&\cF_1
\ar[d]^{\left(\begin{smallmatrix}0&0&1\end{smallmatrix}\right)}
\ar[r]
&\cF_0
\\
\cF_3\ar[r]&\cF_2\ar[r]&\cF_1\ar[r]&\cF_0
} \label{eq:xxexact}
\end{equation}
Since the nonzero entries of (\ref{eq:xxexact}) are sections of $\O$,
they are constants hence holomorphic in the $U_1$ patch as well.  This
also explains that we can't simply multiply (\ref{eq:xxexact}) by $x$
or $x^2$ to conclude exactness of $\fd y \star\fd x$, and $\fd y\star\fd y$,
as these are not holomorphic in the $U_1$ patch.  In fact, it can be
checked that $\fd y\star\fd x$ and $\fd y\star \fd y$ generate
$\Ext^2(\O_C,\O_C)$.

Next, we compute $\fd x\star\fd x\star\fd x$ to be
\begin{equation}
\xymatrix{
&&&\cF_3\ar[r]\ar[d]^{-1}&
\cF_2\ar[r]&
\cF_1\ar[r]&
\cF_0\\
\cF_3\ar[r]&\cF_2\ar[r]&\cF_1\ar[r]&\cF_0
}
\end{equation}
We immediately compute 
\begin{equation}
\fd y\star\fd x\star\fd x=x\left(\fd x\star\fd x\star\fd x\right),\ 
\fd y\star\fd y\star\fd x=x^2\left(\fd x\star\fd x\star\fd x\right),\ 
\fd y\star\fd y\star\fd y=x^3\left(\fd x\star\fd x\star\fd x\right).
\end{equation}

The $\mf{d}$-exactness of $\fd x\star\fd x$ and $\mf{d}$-closedness of
$\fd x$ and $\fd y$ imply that $\fd x\star\fd x\star\fd y$ and $\fd
x\star\fd x\star\fd x$ are $\mf{d}$-exact.  Note that $\fd y\star\fd
y\star\fd y$ is $\mf d$ of
\begin{equation}
\xymatrix@C+8mm@R+6mm{
&&\cF_3\ar[r]\ar[d]^{\left(\begin{smallmatrix}0\\0\\0\end{smallmatrix}
\right)}
&\cF_2
\ar[d]^{\left(\begin{smallmatrix}0&-1&0\end{smallmatrix}\right)}\ar[r]
&\cF_1\ar[r]
&\cF_0
\\
\cF_3\ar[r]&\cF_2\ar[r]&\cF_1\ar[r]&\cF_0
} 
\end{equation}
It can be shown that $\fd y\star\fd y\star\fd x$ is not exact and
generates $\Ext^3(\O_C,\O_C)$.

This shows that $XY^2$ is the only cubic term in the superpotential.
It is not hard to show inductively that all $m_k=f_k=0$ for
$k>2$. Therefore, we have no higher terms in the superpotential and so
\begin{equation}
  W = \Tr(XY^2),
\end{equation}
as might have been expected from \cite{Katz:super}.

%%%%%%%%%%%%%%%%%%%%%%%%%%%%%%%%%%%%%%%%%%%%%%%%%%%%%%%%%%%%%%%%%%%

\section*{Acknowledgments}

We wish to thank M.~Douglas, S.~Kachru, A.~Lawrence, C.~Lazaroiu,
I.~Melnikov, and E.~Sharpe for useful
conversations. P.S.A.~is supported in part by NSF grant
DMS-0301476, Stanford University, SLAC and the Packard Foundation.
S.K.~is supported in part by NSF grants DMS 02-96154, DMS 02-44412,
and NSA grant MDA904-03-1-0050.

%\bibliographystyle{my-phys}
%\bibliography{string}

\end{document}